\documentclass[useAMS,usenatbib]{mn2e}

\usepackage{hyperref}
\usepackage{amsfonts,amsmath,amssymb}
\usepackage{graphicx}
\usepackage{graphics}
\usepackage{epsfig}

\def\baas{Bulletin of the American Astronomical Society}
\def\mnras{Monthly Notices of the Royal Astronomical Society}
\def\na{New Astronomy}
\def\apj{Astroph. J.}
\def\apjl{Astroph. J.}
\def\epjc{European Phys. J. C}
\def\jcap{JCAP}
\def\jkps{J. Korean Phys. Soc.}
\def\ijmpd{Int. J. Mod. Phys. D}

\def\aap{A\&A}
\def\plb{Phys. Lett. B}

\def\prd{Phys. Rev. D}
\def\pre{Phys. Rev. E}
\def\prl{Phys. Rev. Lett.}
\def\RMP{Rev. Mod. Phys.}
\def\pasj{PASJ}

\def\aj{AJ}
\def\ar{Astron. Rep.}
\def\physa{Physica A Statistical Mechanics and its Applications}

\title[Core-Halo distribution of dark matter in galaxies]
  {On the core-halo distribution of dark matter in galaxies}
\author[R. Ruffini et al.]
  {R.~Ruffini,$^{1,2}$\thanks{ruffini@icra.it}
  C.~R.~Arg\"uelles,$^2$ J.~A.~Rueda,$^{1,2}$ \\
  $^1$ Dipartimento di Fisica and ICRA, Sapienza Universit\`a di Roma, P.le Aldo Moro 5, I--00185 Rome, Italy\\
  $^2$ ICRANet, P.zza della Repubblica 10, I--65122 Pescara, Italy
  }
\date{}

\pagerange{}
\pubyear{2015}

\def\LaTeX{L\kern-.36em\raise.3ex\hbox{a}\kern-.15em
    T\kern-.1667em\lower.7ex\hbox{E}\kern-.125emX}

\begin{document}

\label{firstpage}

\maketitle

\begin{abstract}
 We investigate the distribution of dark matter in galaxies by solving the equations of equilibrium of a self-gravitating system of massive fermions (`inos') at selected temperatures and degeneracy parameters within general relativity. Our most general solutions show, as a function of the radius, a segregation of three physical regimes: 1) an inner core of almost constant density governed by degenerate quantum statistics; 2) an intermediate region with a sharply decreasing density distribution followed by an extended plateau, implying quantum corrections; 3) an asymptotic, $\rho\propto r^{-2}$ classical Boltzmann regime fulfilling, as an eigenvalue problem, a fixed value of the flat rotation curves. This eigenvalue problem determines, for each value of the central degeneracy parameter, the mass of the ino as well as the radius and mass of the inner quantum core. Consequences of this alternative approach to the central and halo regions of galaxies, ranging from dwarf to big spirals, for SgrA*, as well as for the existing estimates of the ino mass, are outlined.
\end{abstract}

\begin{keywords}
Methods: numerical -- Cosmology: dark matter -- Galaxies: halos -- Galaxies: nuclei -- Galaxies: structure
\end{keywords}

\section{Introduction}

The problem of identifying the masses and the fundamental interactions of the dark matter particles is currently one of the most fundamental issues in physics and astrophysics. The first astrophysical and cosmological constraints on the mass of the dark matter particle appeared in \cite{1972PhRvL..29..669C,1972gcpa.book.....W,1974ApJ...194..543G,1977PhRvL..39..165L,1979PhRvL..42..407T}. As we will show, some inferences on the dark matter particle mass can be derived from general considerations based solely on quantum statistics and gravitational interactions on galaxy scales.

An important open issue in astrophysics is the description of the dark matter in terms of collisionless massive particles. Attempts have been presented to put constraints on its phase-space density by knowing its evolution from the cosmological decoupling until the approximate time of virialization of a dark matter halo. Phenomenological attempts have been proposed in the past in terms of Maxwellian-like, Fermi-Dirac-like or Bose-Einstein-like distribution functions. Since the 80's all the way up to the present, the problem of modeling the distribution of dark matter in terms of self-gravitating quantum particles has been extensively studied and contrasted against galactic observables. In \cite{1983A&A..119..35R,1993PhLB..306...79V,1998MNRAS..296..569C,2002PrPNP..48..291B,2002PRE..65..5C,2008PRD..77..4B,2013pdmg.conf30204A,2013IJMPD..2260008R,
2013NA..22..39D,2014IJMPD..23..144200A,2014JKPS..65..801A,2014MNRAS..442..2717D,2015arXiv..1402.0695S}, and references therein, this problem was studied by considering Fermi-Dirac statistics in different regimes, from the fully degenerate to the dilute one, and for different fermion masses going from few eV to keV. Instead, in \cite{1994PRD..50..3650S,2000PRL..85..1158H,2007JCAP..6..25B,2008PRD..77..4B,2013MNRAS..428..712S,2014PRD..89..8H} the same problem was analyzed in terms of Bose-Einstein condensates with particle masses from $10^{-25}$~eV up to few eV.

Attempts of studying galactic structures in terms of fundamental physical principles such as thermodynamics and statistical physics, has been long considered (e.g. \cite{2008gady.book..B}) since galaxies present many quasi-universal self-organized properties such as: the constant mean surface density at one-halo scale-length for luminous and dark matter (\cite{2009NATURE..461..627G}); the Fundamental Plane of galaxies (\cite{1987ApJ...313...59D,1996MNRAS.280..167J}); or the fact that dark matter halos can be well fitted by many different but similar profiles that resemble isothermal equilibrium spheres (e.g. \cite{2008AJ....136.2648D,2011AJ....142..109C,2014MNRAS..442..2717D}). Within the statistical and thermodynamical approach, the most subtle problem is the one of understanding the complex processes of relaxation which take place before a galactic halo enters in the steady states we observe. In the context of this paper we will deal only with the (quasi) relaxed states of galaxies, and do not worry about the previous relaxation history of the halos. Nevertheless, and in order to justify in a consistent way the hypothesis we use here, the relaxation process must be certainly considered within the realm of collisionless relaxation, giving the non-interacting nature of the dark matter at halo scales. Formally speaking, this kind of relaxation process differs from the standard collisional relaxation by the fact that the last is described in terms of the Fokker-Planck equation, while the former must be described in terms of the Vlasov-Poisson equation, in order to account for the space and time variations in the overall gravitational potential, not included in the collisional approach (\cite{2008gady.book..B}). While collisional relaxation processes can be applied in globular clusters (stellar component dominant) implying relaxation times $t_R$ of the order or less than the age of the Universe, if applied to galaxies, these processes are largely not relevant because $t_R$ exceeds $10$~Gyr by orders of magnitude (\cite{2008gady.book..B}). By the contrary, it has been extensively shown by now that the time-varying (global) gravitational potential proper of the collisionless process known as violent relaxation (\cite{1967MNRAS.136..101L,2002astro.ph.12205C,2006PhyA..365..102C}), provides a relaxation mechanism analogous to collisions in a gas, but with an associated dynamical time-scale much shorter $t_D<<t_R$; implying now an excellent opportunity to attack the problem of relaxation in galaxies. The central outcome of this theory is that within a few dynamical times $t_D$, the collisionless system quasi relaxes into a tremendously long lived quasi-stationary-state (QSS), which under well mixing conditions can be described in terms of the Fermi-Dirac statistics as shown in \cite{1967MNRAS.136..101L,1978ApJ...225...83S,1996ApJ...466L...1K,2002PRE..65..5C,2002astro.ph.12205C,2005A&A...432..117C,2006PhyA..365..102C}\,\footnote{It has been explicitly shown that these kind of Fermi-Dirac distribution functions can be obtained from a maximization entropy principle at fixed total mass and temperature of the systems (\cite{1998MNRAS..296..569C,2002PRE..65..5C,1999GReGr..31.1105B,2005A&A...432..117C}), implying therefore the necessity for these quasi-relaxed structures to be bounded in radius. This condition can be achieved, for example, by introducing a cut-off in the momentum space of the original Fermi-Dirac distribution as shown first in \cite{1992A&A...258..223I}, and more recently in the context of the model here introduced, in \cite{2013IJMPD..2260008R,2014WS..B..1730F}. The main properties of the fermionic model relevant for the conclusions of this work do not depend on the cut-off as shown in \cite{2014WS..B..1730F}, which only set the outermost boundary radius. Therefore we will adopt for simplicity the standard Fermi-Dirac statistics throughout this paper.}. Even though the Fermi-Dirac distribution was first obtained in terms of a coarse-grained dynamical description (\cite{1967MNRAS.136..101L}), the same statistics was also derived more fundamentally, in terms of particles, either distinguishable (i.e. stars \cite{1978ApJ...225...83S}), or indistinguishable fermionic particles (\cite{1996ApJ...466L...1K,2002PRE..65..5C}), as the ones we are interested here\,\footnote{In any case, the (fermionic) Fermi-Dirac distribution used in this work must be always thought as the final outcome of a macroscopic \textit{coarse-grained mixing}, such that the macroscopic entropy can increase during the complex (collisionless) relaxation processes (second law of thermodynamics) and eventually be maximized to find the final state, as in the cases mentioned in the above footnote 1.}.

Models based on self-gravitating fermions whose equilibrium distributions are assumed to be everywhere in a classical dilute regime (i.e. which can be well approximated by Boltzmannian distributions) as the one recently studied in \cite{2014MNRAS..442..2717D}, may have serious problems of stability when applied to galactic structures such as big spirals. Even though a model of this kind provide good fits when contrasted with observational rotation curves and density profiles (which is also the case within our model, \cite{2015arXiv..1402.0695S}), these profiles most likely undergo core-collapse, being this an inevitable fait of Boltzmannian-based distributions which present large density contrast between center and periphery, even in the case of collisionless particles (\cite{1990PhR...188..285P,1998MNRAS..296..569C,2002astro.ph.12205C}). By the contrary, for self-gravitating systems of collisionless particles which develop some degree of central degeneracy such that the overall dilute-regime can no longer be assumed (i.e. for $\theta_0\gtrsim 10$ within our model), the core-collapse can be stopped, basically because the exclusion principle now present \textit{saturates} the gravitational collapse (see \cite{1998MNRAS..296..569C,1999GReGr..31.1105B,2002PRE..65..5C,2002astro.ph.12205C}).

It is our opinion that in the fermionic case, a clear differentiation of a quantum degenerate core and an almost classical halo, has never been properly implemented. In particular it has been neglected the crucial role of comparing and contrasting different configurations, for fixed halo boundary conditions. As we will show, this leads to a very specific eigenvalue problem for the mass of the inos.

In this paper, and for completeness, we formulate the general problem of the dark matter distribution in galaxies based in the following assumptions: 1) that the dark matter phase-space density is described by the Fermi-Dirac statistics; 2) that the equilibrium equations for the configurations be solved within a general relativistic treatment; 3) we set the boundary condition for all dark matter profiles associated with a specific galaxy type (dwarfs, spirals, and big spirals), to have, in each case, the same value of the flat rotation curve. Having established this procedure in section \ref{sec:2}, we evidence in section \ref{sec:3}: i) the new core-halo distribution of dark matter density, which is composed by a dense compact core governed by almost degenerate quantum statistics, a semi-degenerate transition, followed by a dilute halo governed by Boltzmann classic statistics; ii) for each central degeneracy parameter we determine as an eigenvalue problem, the mass and radius of the inner quantum core, as well as the corresponding ino mass; and iii) we show that, for an ino mass of $\sim 10$~keV/$c^2$, there is in our model a theoretical correlation between the inner quantum core mass and the halo mass, for galaxy types from dwarf up to big spirals. From these considerations clearly follows that the determination of the ino mass is uniquely established by the properties of the inner quantum core and the asymptotic boundary conditions, and it cannot be determined in a dark matter distribution governed only by a Boltzmannian distribution, which is independent of the mass of the ino. In section \ref{sec:4} we summarize and discuss our results.

\section{Equilibrium equations and boundary conditions}\label{sec:2}

Following \cite{1990A&A...235....1G,2014JKPS..65..801A}, we here consider a system of general relativistic
self-gravitating bare massive fermions under the approximation of thermodynamic equilibrium. As mentioned above, this approximation is well justified under the assumption of well mixing during the collisionless relaxation process, where the overall distribution function of the inos in the QSS, can be well approximated by the Fermi-Dirac distribution. No additional
interactions are initially assumed for the fermions besides their fulfillment of quantum-like statistics and the relativistic gravitational equations. In particular, we do not assume weakly interacting particles as in \cite{1979PhRvL..42..407T}. We refer to this \emph{bare} particles more generally as \emph{inos}, leaving the possibility of additional fundamental interactions to be determined by further requirements to be fulfilled by the model. Already this treatment of bare fermions leads to a new class of equilibrium configurations and, correspondingly, to new limits to the ino mass. This is a necessary first step in view of a final treatment involving additional interactions to be treated self-consistently, as we will soon indicate here.

The density and pressure of the fermion system are given by
%
\begin{align}
    \rho &= m\frac{2}{h^3}\int f(p)\left[1+\frac{\epsilon(p)}{m c^2}\right]\,d^3p,\label{eq:rho}\\
    P &= \frac13 \frac{2}{h^3}\int
    f(p)\left[1+\frac{\epsilon(p)}{m c^2}\right]^{-1}\left[1+\frac{\epsilon(p)}{2 m c^2}\right]\epsilon\,d^3p,\label{eq:p}
\end{align}
where the integration is over all the momentum space, $f_p=(\exp[(\epsilon-\mu)/(k T)]+1)^{-1}$ is the distribution function, $\epsilon=\sqrt{c^2 p^2+m^2 c^4}-mc^2$ is the particle kinetic energy, $\mu$ is the chemical potential with the particle rest-energy subtracted off, $T$ is the temperature, $k$ is the Boltzmann constant, $h$ is the Planck constant, $c$ is the speed of light, and $m$ is the ino's particle mass. We do not include the presence of anti-fermions, i.e.~we consider temperatures $T \ll m c^2/k$.

The Einstein equations for the spherically symmetric metric $g_{\mu \nu}={\rm diag}(e^{\nu},-e^{\lambda},-r^2,-r^2\sin^2\Theta)$, being $\Theta$ the azimutal angle, where $\nu$ and $\lambda$ depend only on the radial coordinate $r$, together with the thermodynamic equilibrium conditions of \cite{1930PhRv...35..904T}, $e^{\nu/2} T=$constant, and \cite{klein49}, $e^{\nu/2}(\mu+m c^2)=$constant,
%
%
can be written as \cite{1990A&A...235....1G}
\begin{align}
		\frac{d\hat M}{d\hat r}&=4\pi\hat r^2\hat\rho, \label{eq:eqs1}\\
		\frac{d\theta}{d\hat r}&=-\frac{1-\beta_0(\theta-\theta_0)}{\beta_0}
    \frac{\hat M+4\pi\hat P\hat r^3}{\hat r^2(1-2\hat M/\hat r)},\label{eq:eqs2}\\
    \frac{d\nu}{d\hat r}&=\frac{2(\hat M+4\pi\hat P\hat r^3)}{\hat r^2(1-2\hat M/\hat r)}, \\
    \beta_0&=\beta(r) e^{\frac{\nu(r)-\nu_0}{2}}\, . \label{eq:tolman2}
\end{align}
%
The following dimensionless quantities were introduced: $\hat r=r/\chi$, $\hat M=G M/(c^2\chi)$, $\hat\rho=G \chi^2\rho/c^2$, $\hat P=G \chi^2 P/c^4$,
where $\chi=2\pi^{3/2}(\hbar/mc)(m_p/m)$, with $m_p=\sqrt{\hbar c/G}$ the Planck mass, and the temperature and degeneracy parameters, $\beta=k T/(m c^2)$ and $\theta=\mu/(k T)$, respectively. The constants of the Tolman and Klein conditions are evaluated at the center $r=0$, indicated with a subscript `0'.

The system variables are $[M(r),\theta(r),\beta(r),\nu(r)]$. We integrate Eqs.~(\ref{eq:eqs1}--\ref{eq:tolman2}) for given initial conditions at the center, $r=0$, in order to be consistent with the observed dark matter halo mass $M(r=r_h)=M_h$ and radius $r_h$, defined in our model at the onset of the flat rotation curves. The so called \textit{halo radius} (and mass) in this paper represent the one-halo scale length (and mass) associated with the fermionic model here presented, and corresponding with the turn-over of the density profiles in total analogy as other halo-scale lengths used in the literature such as $r_0$ or $r_{-2}$ as shown in Fig.~\ref{fig:OursvsNFW}. The circular velocity is
%
\begin{equation}\label{eq:vc}
v(r)=\sqrt{\frac{G M(r)}{r-2 G M(r)/c^2}}\, ,
\end{equation}
which at $r=r_h$, is $v(r=r_h)=v_h$.

It is interesting that a very similar set of equations have been re-derived in \cite{2002PrPNP..48..291B} apparently disregarding the theoretical approach already implemented in 1990 in \cite{1990A&A...235....1G}. They integrated the Einstein equations fixing a fiducial mass of the ino of $m=15$~keV$/c^2$, and they derived a family of density profiles for different values of the central degeneracy parameter at a fixed temperature consistent with an asymptotic circular velocity $v_{\infty}=220$~km/s. They conclude that a self-gravitating system of such inos could offer an alternative to the interpretation of the massive black hole in the core of SgrA* (\cite{2008ApJ...689.1044G}). Although this result was possible at that time, it has been superseded by new constraints imposed by further observational limits on the trajectory of S-stars such as S1 and S2 (\cite{2008ApJ...689.1044G,2009ApJ...707L.114G}).

In this paper we give special attention to the halo boundary conditions determined through the flat rotation curves. We integrate our system of equations using different boundary conditions to the ones imposed in \cite{2002PrPNP..48..291B} and reaching different conclusions. We first apply this model to typical spiral galaxies, similar to our own galaxy, adopting dark matter halo parameters (\cite{2008AJ....136.2648D,2009PASJ...61..227S}):
\begin{equation}
r_h=25~{\rm kpc},\,\, v_h=168~{\rm km/s},\,\, M_h=1.6\times10^{11} M_\odot\, .
\label{eq:observablesI}
\end{equation}

\begin{figure*}
\centering
\includegraphics[width=.31\hsize,clip]{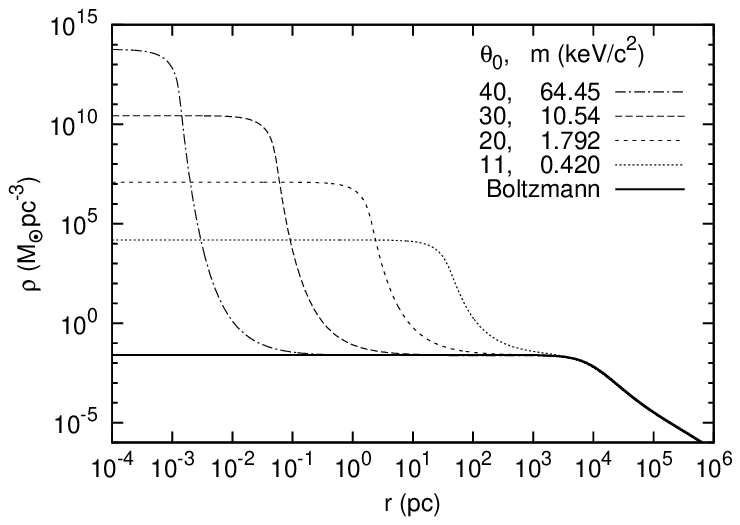}
\includegraphics[width=.31\hsize,clip]{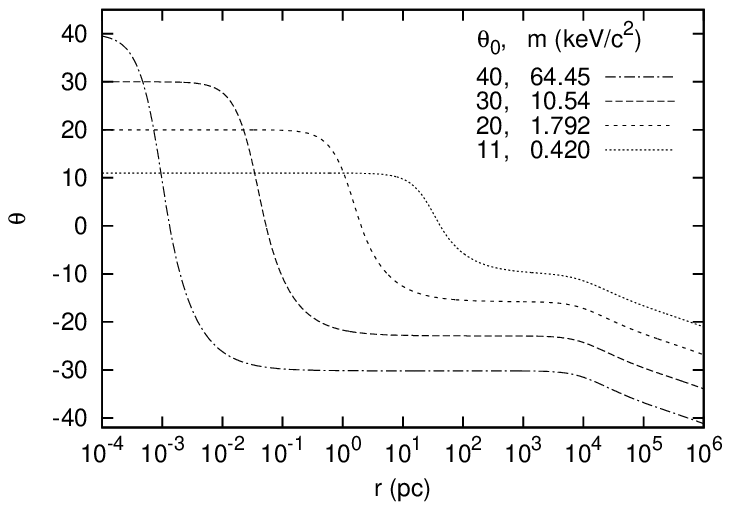}
\includegraphics[width=.31\hsize,clip]{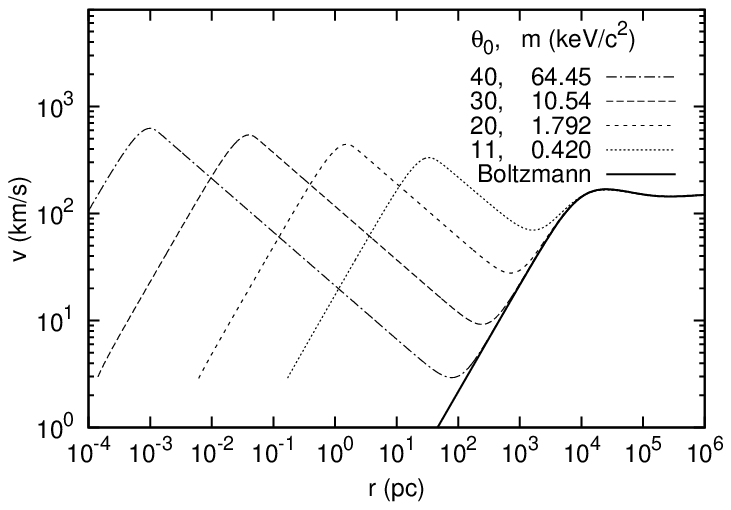}
\caption{Mass density (left panel), degeneracy parameter (central panel), and rotation velocity curves (right panel) for specific ino masses $m$ and central degeneracies $\theta_0$ fulfilling the observational constraints (\ref{eq:observablesI}). The density solutions are contrasted with a Boltzmannian isothermal sphere with the same halo properties. All the configurations, for any value of $\theta_0$ and corresponding $m$, converge for $r\gtrsim r_h$ to the classical Boltzmannian isothermal distribution. It is clear how the Boltzmann distribution, is as it should be, independent of $m$. Interestingly, when the value $M_c(r\lesssim10^{-2}$ pc)$\sim 10^6 M_\odot$ (i.e. $m\sim 10$~keV$/c^2$) is chosen as the one of more astrophysical interest, the onset of the classical Boltzmann regime takes place at distances of $r\gtrsim$ few $10^2$~pc, in consistency with the observed cored nature of the innermost resolved regions in spiral galaxies as analyzed in \citep{2008AJ....136.2648D}.}
\label{fig:rhothfamilies}
\end{figure*}

Later on we repeat the analysis also for typical dwarf spheroidal galaxies: $r_h=0.6$~kpc; $v_h=13$~km/s; $M_h=2\times10^7 M_\odot$ (\cite{2009ApJ...704.1274W}); as well as for typical big spiral galaxies:$r_h=75$~kpc; $v_h=345$~km/s; $M_h=2\times10^{12} M_\odot$ (\cite{2009arXiv0911.1774B}). The initial conditions are $M(0)=0$, $\nu(0)=0$, $\theta(0)=\theta_0$ and $\beta(0)=\beta_0$. We integrate Eqs.~(\ref{eq:eqs1}--\ref{eq:tolman2}) for selected values of $\theta_0$ and $m$, corresponding to different degenerate states of the gas at the center of the configuration. The value of $\beta_0$ is actually an eigenvalue which is found by a trial and error procedure until the observed values of $v_h$ and $M_h$ at $r_h$ are obtained. We show in Fig.~\ref{fig:rhothfamilies} the density profiles and the rotation curves as a function of the distance for a wide range of parameters ($\theta_0,m$), for which the boundary conditions in (\ref{eq:observablesI}) are exactly fulfilled.

\section{Dark matter profiles: from dwarf to big spiral galaxies}\label{sec:3}

\begin{figure}
\centering
\includegraphics[width=\hsize,clip]{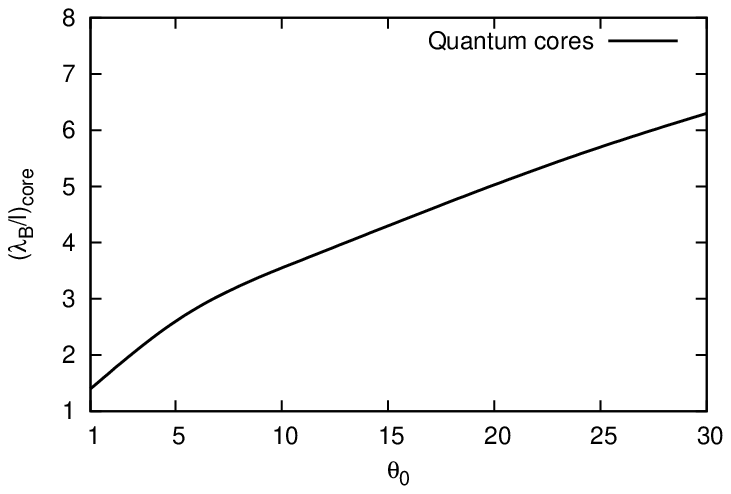}
\caption{
The less degenerate quantum cores in agreement with the halo observables (\ref{eq:observablesI}) corresponds to $\theta_0\approx 10$ ($\lambda_B\sim3l_c$). These cores are the ones which achieve the largest sizes, of order $\sim 10^1$~pc, and implying the lowest ino masses in the sub-keV region.}
\label{fig:coredeg}
\end{figure}

The phase-space distribution encompasses both the classical and quantum regimes. Correspondingly, the integration of the equilibrium equations leads to three marked different regimes (see Fig.~\ref{fig:rhothfamilies}): a) the first consisting in a quantum core of almost degenerate fermions. These cores are characterized by having $\theta(r)>0$. The core radius $r_c$ is defined by the first maximum of the velocity curve. A necessary condition for the validity of this quantum treatment for the central core is that the interparticle mean-distance, $l_c$, be smaller or of the same order, of the thermal de Broglie wavelength of the inos, $\lambda_B=h/\sqrt{2\pi m kT}$. As we show below (see Fig.~\ref{fig:coredeg}), this indeed is fulfilled in all the cases here studied. b) A second regime where $\theta(r)$ goes from positive to negative values for $r>r_c$, all the way up to the so called classical domain where the quantum corrections become negligible. This transition region consists in a sharply decreasing density followed by an extended plateau. c) The classical regime described by Boltzmann statistics and corresponding with $\theta(r)\ll-1$ (for $r\gtrsim r_h$), in which the solution tends to the Newtonian isothermal sphere with $\rho\sim r^{-2}$, where the flat rotation curve sets in. Of course, the flat region of the velocity curve can not continue indefinitely in the case of realistic bounded systems. This can be easily achieved in the context of our model and without changing the results here presented, by introducing a cut-off in the momentum space accounting for possible dissipative and/or tidal effects as done in (\cite{2013IJMPD..2260008R,2014WS..B..1730F}), and already explained in the footnote 1. Regarding a possible astrophysical discussion about the novel (increasing-decreasing) aspect of the inner part of the rotation curve arising before reaching the known classical behaviour, as well as the numerical implications of $\beta_0$ and $\theta_0$, they are given at the end of this section.

We define the core mass, the circular velocity at $r_c$, and the core degeneracy as $M_c=M(r_c)$, $v_c=v(r_c)$ and $\theta_c=\theta(r_c)$, respectively. In Table \ref{table:1} we show the core properties of the equilibrium configurations in spiral galaxies, for a wide range of ($\theta_0,m$). For any selected value of $\theta_0$ we obtain the correspondent ino mass $m$ to fulfill the halo properties (\ref{eq:observablesI}), after the above eigenvalue problem of $\beta_0$ is solved.

\begin{table*}
\centering
\begin{tabular}{@{}|c|c|c|c|c|c|@{}}
\hline
$\theta_0$ & $m$~(keV/c$^2$)  & $r_c$~(pc) & $M_c (M_\odot)$ & $v_c$~(km/s) & $\theta_c$\\
\hline

11 & 0.420 & $3.3\times10^1$ &  $8.5\times10^8$ & $3.3\times10^2$ & $2.1$\\
25 & 4.323 & $2.5\times10^{-1}$ &  $1.4\times10^7$ & $4.9\times10^2$ & $5.5$\\
30 & 10.540 & $4.0\times10^{-2}$ & $2.7\times10^6$ & $5.4\times10^2$ & $6.7$\\
40 & 64.450 & $1.0\times10^{-3}$ & $8.9\times10^4$ & $6.2\times10^2$ & $8.9$\\
58.4 & $2.0\times10^3$ & $9.3\times10^{-7}$ & $1.2\times10^2$ & $7.5\times10^2$ & $14.4$\\
98.5 & $3.2\times10^6$ & $3.2\times10^{-13}$ & $7.2\times10^{-5}$ & $9.8\times10^2$ & $21.4$\\
\hline
\end{tabular}
\caption{Core properties for different equilibrium configurations fulfilling the halo parameters (\ref{eq:observablesI}) of spiral galaxies.}
\label{table:1}
\end{table*}

It is clear from Table \ref{table:1} and Fig.~\ref{fig:rhothfamilies} that the mass of the core $M_c$ is strongly dependent on the ino mass, and that the maximum space-density in the core is considerably larger than the maximum value considered in \citep{1979PhRvL..42..407T} for a Maxwellian distribution.
Interestingly, as can be seen from Fig.~\ref{fig:rhothfamilies}, the less degenerate quantum cores in agreement with the halo observables (\ref{eq:observablesI}), are the ones with the largest sizes, of the order of halo-distance-scales. In this limit, the fermion mass acquires a sub-keV minimum value which is larger, but comparable, than the corresponding sub-keV bound in \citep{1979PhRvL..42..407T}, for the same halo observables. Indeed, their formula gives a lower limit $m\approx 0.05$~keV$/c^2$ when using the proper value for the King radius, $r_K\simeq8.5$~kpc, as obtained from $\sigma=\sqrt{2/5}v_h$ and $\rho_0=2.5\times10^{-2} M_\odot/$pc$^3$, which are the associated values to the Boltzmannian density profile of Fig.~\ref{fig:rhothfamilies}. This small difference is formally understood by the following fact: while their conclusions are reached by adopting the maximum phase-space density, $Q_{max}^h\sim\rho^h_0 m^{-4}\sigma_h^{-3}$, at the center of a halo described by a Maxwellian distribution; in our model the maximum phase-space density is reached at the center of the dense quantum core described by Fermi-Dirac statistics, $Q_{max}^c\sim\rho^c_0 m^{-4}\sigma_c^{-3}$ (where lower and upper index $c$ reads for the central core). An entire new family of solutions exists for larger values of central phase-space occupation numbers, always in agreement with the halo observables (see Fig.~\ref{fig:rhothfamilies}). Now, since these phase-space values, by the Liouville's theorem, can never exceed the maximum primordial phase-space density at decoupling, $Q_{max}^d$, we have $Q_{max}^{h,c}<Q_{max}^d$. Then, considering that all our quantum solutions satisfy $Q_{max}^c>Q_{max}^h$, it directly implies larger values of our ino mass with respect to the Tremaine and Gunn limit. Nevertheless, as we have quantitatively shown above, e.g.~for the case of typical spiral galaxies, the two limits become comparable for our less degenerate ($\theta_0\approx 10$) quantum cores in agreement with the used halo observables (\ref{eq:observablesI}).

In the case of a typical spiral galaxy, for an ino mass of $m\sim10$~keV$/c^2$, and a temperature parameter $\beta_0\sim10^{-7}$, obtained from the observed halo rotation velocity $v_h$, the de Broglie wavelength $\lambda_B$ is higher than the interparticle mean-distance in the core $l_c$, see Fig.~\ref{fig:coredeg}, safely justifying the quantum-statistical treatment applied here.

If we turn to the issue of an alternative interpretation to the black hole on SgrA*, we conclude that a compact degenerate core mass $M_c\sim4\times 10^6 M_\odot$ is definitely possible corresponding to an ino of $m\sim10$~keV$/c^2$ (see Table \ref{table:1}). However, the core radius of our configuration is larger by a factor $\sim 10^2$ than the one obtained with the closest observed star to Sgr A*, i.e. the S2 star (\cite{2009ApJ...707L.114G}).
Nevertheless, for an ino mass of $m\sim10$~keV$/c^2$ ($\theta_0=30$), the very low temperature of the dense quantum core is already a small fraction of the Fermi energy (i.e. $\lambda_B>l$), where additional interactions between the inos should arise, affecting the mass and radius of the new denser core depending on the interaction adopted \footnote{This is analogous for instance to the case of neutron stars, where nuclear fermion interactions strongly influence the mass-radius relation (see, e.g., \cite{2007PhR...442..109L})}. Indeed, we have recently applied this novel idea in \cite{companion}, achieving now higher possible compactness for the new quantum core, in perfect agreement with the observational constraints imposed by the S2 star, and always for ino masses in the range of $m\sim 10^1$~keV$/c^2$. Moreover, the relevance of self-interactions in ultra-cold fermionic-particle collisions has been already shown in laboratory, for example, for (effective) Fermi gases, e.g.~$^6$Li, at temperatures of fractions of the Fermi energy (\cite{2008RvMP...80.1215G}). There, a good agreement between experiment and theory was achieved for such a cold Fermi gas when studied in terms of a grand-canonical many-body Hamiltonian in second quantization, with a term accounting for fermion-fermion interaction, similarly as done in \cite{companion}.

\begin{figure}
\centering
\includegraphics[width=\hsize,clip]{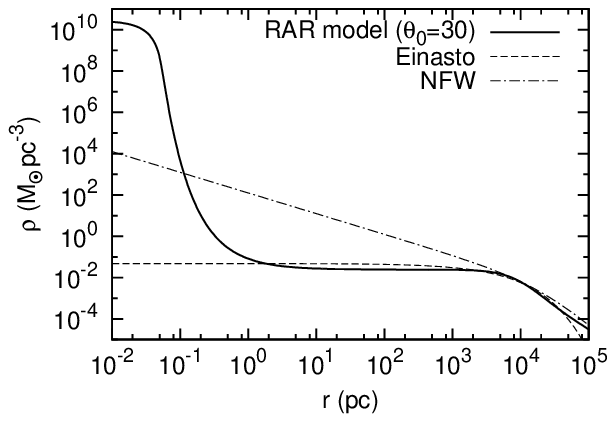}
\caption{The cored behavior of the dark matter density profile from the Ruffini-Arg\"uelles-Rueda (RAR) model is contrasted with the cuspy Navarro-Frenk-White (NFW) density profile \citep{1997ApJ...490..493N}, and with a cored-like Einasto profile \citep{1965TrAlm...5...87E,1989A&A...223...89E}. The free parameters of the RAR model are fixed as $\beta_0=1.251\times10^{-7}$, $\theta_0=30$ and $m=10.54$~keV$/c^2$. The corresponding free parameters in the NFW formula $\rho_{NFW}(r)=\rho_0 r_0/[r(1+r/r_0)^2]$ are chosen as $\rho_0=5\times10^{-3} M_\odot$ pc$^{-3}$ and $r_0=25$~kpc, and for the Einasto profile $\rho_E(r)=\rho_{-2}\exp{[-2n(r/r_{-2})^{1/n}-1]}$, $\rho_{-2}=2.4\times10^{-3} M_\odot$ pc$^{-3}$, $r_{-2}=16.8$~kpc, and $n=3/2$. In the last two models, the chosen free parameters are typical of spiral galaxies according to \citep{2008AJ....136.2648D,2011AJ....142..109C}.}\label{fig:OursvsNFW}
\end{figure}

We further compare and contrast in Fig.~\ref{fig:OursvsNFW} our theoretical curves of Fig.~\ref{fig:rhothfamilies} with observationally inferred ones. In order to provide a more detailed comparison, we have extensively contrasted our three-parametric fermionic model with many other dark matter parametric models resulting from N-body simulations, in terms of a formal Bayesian statistical analysis and using high resolution data samples including for baryonic components, in \cite{2015arXiv..1402.0695S}. It is interesting that the quantum statistical treatment (including relativistic effects) considered here, is characterized by the presence of central cored structures unlike the typical \emph{cuspy} configurations obtained from a classic non-relativistic approximation, such as the ones of numerical N-body simulations in \cite{1997ApJ...490..493N}. This naturally leads to a first step, in terms of a first principle physics approach, to understand the well-known core-cusp discrepancy as first shown in \cite{2001AJ....122.2396D} and further confirmed for typical spiral galaxies in \cite{2011AJ....142..109C}.
Such a difference between the ino's core and the cuspy NFW profile, as well as the possible black hole nature of the compact source in SgrA*, will certainly reactivate the development of observational campaigns in the near future. There the interesting possibility, in view of the BlackHoleCam Project based on the largest Very Long Baseline Interferometry (VLBI) array\footnote{http://horizon-magazine.eu/space}, to verify the general relativistic effects expected in the surroundings of the central compact source in SgrA*. Such effects depend on whether the source is modeled in terms of the RAR model presented here (with the possible inclusion of fermion interactions, \cite{companion}), or as a black hole. To compare and contrast these two alternatives is an observational challenge now clearly open.

Following the analysis developed here for a typical spiral, we have also considered two new different sets of physical dark matter halos: $r_h=0.6$~kpc; $v_h=13$~km/s; $M_h=2\times10^7 M_\odot$ for typical dwarf spheroidal galaxies, (e.g.~\cite{2009ApJ...704.1274W}); and $r_h=75$~kpc; $v_h=345$~km/s; $M_h=2\times10^{12} M_\odot$ for big spiral galaxies, as analyzed in \cite{2009arXiv0911.1774B}.
For big spirals, $\lambda_B/l_c=5.3$, while for typical dwarfs galaxies $\lambda_B/l_c=4.1$, justifying the quantum treatment in both cases.

A remarkable outcome of the application of our model to such a wide range of representative dark halo galaxy types, from dwarfs to big spirals, is that \emph{for the same} ino mass, $m\sim10$~keV/$c^2$, we obtain respectively core masses $M_c\sim10^4 M_\odot$ and radii $r_c\sim10^{-1}$pc for dwarf galaxies, and core masses $M_c\sim 10^7 M_\odot$ and radii $r_c\sim10^{-2}$pc for big spirals. This leads to a possible alternative to intermediate ($\sim10^4 M_\odot$) and more massive ($\sim10^{6-7} M_\odot$) black holes, thought to be hosted at the center of the galaxies.

Moreover, we have obtained out of first principles, a possible universal relation between the dark matter halos and the super massive dark central objects. For a fixed ino mass $m=10$~keV/$c^2$, we found the $M_c$-$M_h$ correlation law
\begin{equation}
\frac{M_c}{10^6 M_\odot}=2.35\left(\frac{M_h}{10^{11} M_\odot}\right)^{0.52},
\label{eq:McMhcorrel}
\end{equation}
%
valid for core masses $\sim [10^4,10^7]\,M_\odot$ (corresponding to dark matter halo masses $\sim[10^7,10^{12}]\,M_\odot$). Regarding the observational relation between massive dark compact objects and bulge dispersion velocities in galaxies (the $M_c$-$\sigma$ relation (\cite{2002chee.conf....3F}), it can be combined with two observationally inferred relations such as the $\sigma$-$V_c$ and the $V_c$-$M_h$ correlations, where $V_c$ is the observed halo circular velocity and $M_h$ a typical halo mass. This was done in \cite{2002ApJ...578...90F} to find, by transitivity, a new correlation between central mass concentrations and halo dark masses ($M_c$-$M_h$). Interestingly, such a correlation matches with the one found above in Eq.~(\ref{eq:McMhcorrel}) in the range $M_c=[10^6,10^7]\,M_\odot$, without assuming the black hole hypothesis. The appearance of a core surrounded by a non-relativistic halo, is a key feature of the configurations presented in this paper. It cannot however be extended to quantum cores with masses of $\sim10^{9} M_\odot$. Such core masses, observed in Active Galactic Nuclei (AGN), overcome the critical mass value for gravitational collapse $M_{cr}\sim M_{pl}^3/m^2$ for keV-fermions, and therefore these cores have to be necessarily black holes (\cite{2014IJMPD..23..144200A}). The characteristic signatures of such supermassive black-holes, including jets and X-ray emissions, are indeed missing from the observations of the much quiet SgrA* source, or the centers of dwarf galaxies.

At this point it is relevant to discuss the qualitative and quantitative relevance of the general relativistic approach proposed here to model the distribution of dark matter in galaxies, when compared with a classical Newtonian approach. For the example analyzed here, i.e. for $m\sim 10$~keV/$c^2$ and spiral galaxies, the compactness of the quantum core is $G M_c/(r_c c^2)\sim 10^{-6}$, thus general relativistic effects are not dominant in these configurations. Under those conditions, we do expect a Newtonian approach to describe satisfactorily the configurations. Indeed, by integrating the corresponding equations of equilibrium in the Newtonian case, which are obtained in the non-relativistic weak-field limit of the treatment presented here \footnote{This is obtained by taking the limit $c\rightarrow\infty$ and $e^{\nu/2}\approx 1+\phi/c^2$, leading to thermodynamic equilibrium conditions $T=$constant, and $\mu+m\phi=$constant, with $\phi$ the Newtonian gravitational potential.}, we obtain similar results to the general relativistic solution within 1\% (for spiral galaxies and $m\sim 10$~keV/$c^2$), keeping the core-halo structure containing the three markedly different physical regimes from quasi-degeneracy regime in the core all the way up to Boltzmaniann one in the halo. As we have explained, such a change of regimes is due to the combination of the non-zero temperature and the changing fermion chemical potential with distance, which produces a changing degeneracy parameter with the distance. It is important to mention at this point that, if we were to model the galactic halos assuming a zero temperature, we would obtain a different behavior of the density profile (resembling our quantum core and never reaching the plateau plus Boltzmannian phase) which leads to non-flat rotation velocity curves (with a raising part, a maximum, and a Keplerian falling down region), hence inconsistent with observations.

A general relativistic treatment becomes a necessity in the case of more compact configurations approaching the critical mass for gravitational collapse, $M_{cr}\sim M_{pl}^3/m^2\sim 10^9 M_\odot$, which as we have shown (\cite{2014IJMPD..23..144200A}) could be attained in the central compact cores observed in AGNs by the same dark matter candidate of $m\sim 10$~keV/$c^2$, and corresponding to different boundary conditions as contrasted with the case of normal galaxies here considered.

We turn now to briefly discuss the astrophysical implications of the full morphology of the dark matter rotation curves as well as the numerical implications of the typical temperature and degeneracy parameters found here. Indeed, the issue addressed in the present article is referred only to a pure dark matter composition while the observational data refers to the sum of the dark and baryonic (gas and stellar populations) matter. The key result presented here is that the dark matter contribution is always predominant in the inner core (at sub-pc scales), and in the outer halo region at the onset of the flat part of the given rotation curve; while in between baryonic matter prevails. We can see from the right panel of Fig.~\ref{fig:rhothfamilies} that indeed, for a Milky Way-like galaxy, our model correctly predicts both the value and flattening of the circular velocity at distances $r\gtrsim 10$~kpc. A detailed comparison of the theoretical curves analyzed here with extended astrophysical data will be soon presented elsewhere (\cite{companion2}), including the special behavior of the circular velocity e.g. at the sub-pc scales.

Regarding the actual values of the dark matter temperatures and (effective) chemical potentials obtained out of the free parameters of the model ($\beta_0$, $\theta_0$) consistent with the (quasi) relaxed galactic structures analyzed here, we have: for an ino mass of $m\sim 10$~keV/$c^2$, in typical dwarfs $T_d\sim 10^{-1}$~K ($\beta_0\sim 10^{-9}$), while in spirals $T_s\sim 10^1$~K ($\beta_0\sim 10^{-7}$). This values when combined with the central degeneracy parameters gives, for typical dwarfs $\mu_{d}\sim 10^{-7}$~keV ($\theta_{0,d}=15$), and $\mu_{s}\sim 10^{-5}$~keV ($\theta_{0,s}=30$) or $\mu_{bs}\sim 10^{-4}$~keV ($\theta_{0,bs}=36$), for typical spiral or big spirals respectively\footnote{It is important to recall that i) due to the small general relativistic effects in the cases analyzed in this work, from the Tolman and Klein conditions the central values of $T$ and $\mu$ given above are accurate through the overall configurations; and ii) $\mu$ is the chemical potential with the fermion rest-mass subtracted-off, therefore the (effective) chemical potentials, including the fermion rest-mass, are roughly (for all the cases with $m \sim 10$~keV$/c^2$) $\mu_{d,s,bs}+m c^2 \approx 10$~keV.}. The issue of the potential implications of these dark matter temperatures and chemical potentials in relation with different possible microscopic models for the dark matter candidate in cosmology, will be a subject for future works.

\section{Conclusions}\label{sec:4}

A consistent treatment of self-gravitating fermions within general relativity has been here introduced and solved with standard boundary conditions appropriate to flat rotation curves observed in galactic halos of spiral and dwarf galaxies. A new structure has been identified: 1) a core governed by quantum-like statistics; 2) a velocity of rotation at the surface of this core which is bounded independently of the mass of the particle and remarkably close to the asymptotic rotation curve; 3) a semi-degenerate region leading to an asymptotic regime described by a pure Boltzmann distribution, consistent with the flat rotation curves observed in galaxies. Interestingly it has been recently shown that quasi relaxed core-halo structures analogous as the one obtained here for the dark matter in galaxies, take part of a broader and more ubiquitous behaviour in nature, proper of long-range collisionless interacting systems including also plasmas and kinetic spin models (\cite{2014PhR...535....1L}).

For $m\sim10$~keV/$c^2$ a universal relation between the mass of the core $M_c$ and the mass of the halo $M_h$ has been found. This universal relation applies in a vast region of galactic systems, ranging from dwarf to big spiral galaxies with core masses $\sim [10^4,10^7]\,M_\odot$ (corresponding to dark matter halo masses $\sim [10^7,10^{12}]\,M_\odot$).

Starting from the basic treatment here introduced, of bare self-gravitating fermions, we have already examined the possibility to introduce new types of interactions (\cite{companion}) among the inos, considering, for example, a self-interacting picture in the context of right-handed sterile neutrinos in the minimal standard model extension (see e.g.~\cite{2009ARNPS..59..191B}), as a viable candidate for the ino particles in our new scenario. The extended approach studied in \cite{companion} allowed us to verify the possibility of the radius of the quantum core to become consistent with the observations of SgrA*, and so open the way to identify additional (effective) fundamental interactions in the ino physics. For this more general analysis, as well as for the model extension which allow us to deal with the very massive galactic compact cores of $M_c\sim10^9 M_\odot$ as studied in \cite{2014IJMPD..23..144200A}, the General Relativistic treatment here introduced for completeness, clearly becomes mandatory.

After this generalized treatment, we will further address the issue of the implications of these kev-fermions in cosmology.\\\\

\textbf{Acknowledgments}

This work was supported by the International Center for Relativistic Astrophysics Network (ICRANet). The authors want give a special thank to the anonymous referee for the clever suggestions regarding the astrophysical implications of the model here presented.



\end{document}